\documentclass[prb,twocolumn,preprintnumbers,amsmath,amssymb,superscriptaddress]{revtex4}
\usepackage{bm}
\usepackage{amsmath}
\usepackage{dcolumn}
\usepackage[dvips]{graphicx}
\begin{document}
\bibliographystyle{apsrev}

\title{Theory of Anisotropic Hopping Transport  due to Spiral Correlations  in
the Spin-Glass Phase of Underdoped Cuprates}
\author{Valeri N. Kotov}
\email{valeri.kotov@epfl.ch}
\affiliation{Institute of Theoretical Physics,
 Swiss Federal Institute of Technology (EPFL), \\
CH-1015 Lausanne, Switzerland}
\author{Oleg P. Sushkov}
\email{sushkov@phys.unsw.edu.au}
\affiliation{School of Physics, University of New South Wales, Sydney 2052, Australia}

%\date{\today}
\begin{abstract}
We study the in-plane resistivity  anisotropy in the spin-glass
 phase of the high-$T_{c}$ cuprates, on the basis of  holes moving  in a spiral spin background.
This picture follows from analysis of the extended $t-J$ model with
Coulomb impurities.
In the variable-range hopping regime the resistivity anisotropy is found to
 have a maximum value of around 90\%, and it decreases with temperature,
 in excellent agreement with experiments in La$_{2-x}$Sr$_x$CuO$_4$.  
In our approach the transport anisotropy is due to the
 non-collinearity of the spiral spin state, 
rather than an intrinsic tendency of the charges to self-organize.

\end{abstract}
%\pacs{74.72.Dn, 75.10.Jm, 75.30.Fv, 75.50.Ee}
\maketitle

\section{Introduction}
%{\it Introduction.}
 One of the central issues in the physics of the
 high-temperature superconductors  is  the nature
 of the ground state at low doping. In particular,
 the possible  co-existence of ordering tendencies
 in the spin and charge sectors at low temperature  
is currently being actively investigated. \cite{Orenstein,Sachdev,Kivelson1}
%as such types of order  reflect the fundamental properties of holes doped in antiferromagnets.
Experimentally  at low temperature   
 La$_{2-x}$Sr$_x$CuO$_4$ (LSCO) co-doped with Nd (LNSCO) 
exhibits  static lattice deformation \cite{Tranquada} as well
 as incommensurate (IC)  magnetic order  at doping $x \approx 1/8$.
The lattice deformation indicates the presence of static charge order (stripes).
Recently similar behavior has also been found in
 La$_{2-x}$Ba$_x$CuO$_4$ (LBCO), \cite{LBCO} $x \approx 1/8$. 
However for  $x < 1/8$ and in Nd-free LSCO, charge  order 
has not been detected, while
 IC magnetism persists down to $x\approx0.02$. \cite{Wakimoto, Fujita}
Thus  while IC magnetic order is generically observed in the
 underdoped regime,  static charge order seems to be confined to
 the neighborhood of $x = 1/8$.

\vspace{0.2cm}
\begin{figure}[ht]
\centering
\includegraphics[height=115pt,keepaspectratio=true]{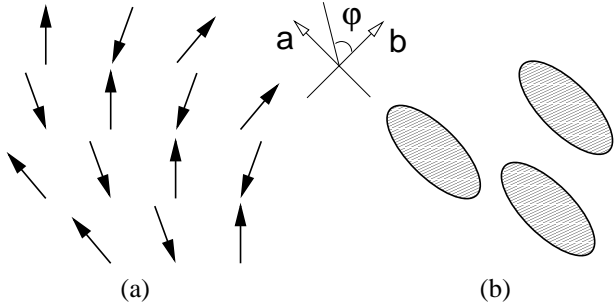}
\caption{(a) (1,1) Spiral spin state. (b) Localized (Coulomb trapped) holes, with positions
 pinned by the random Sr impurities (schematic). The ellipses reflect 
 the symmetry of the wave-function (\ref{solution}). 
The orthorhombic $\hat{a}$ and $\hat{b}$ directions are shown.}
\label{Fig1}
\end{figure}
\noindent

Theoretically the  IC magnetism in the cuprates 
is often modeled as originating from static
 charge stripes, and since these are not universally
 present, one is forced to introduce additional
 concepts, such as ``fluctuating" stripes, electronic
 liquid crystals, etc. \cite{Kivelson1,Kivelson2}
 The  microscopic origin of  such states is still not clear and is currently being
 debated. An alternative explanation for the IC magnetism which  
follows naturally from the $t-J$ model,  is the formation of a non-collinear spiral state,  
 partially  relieving the frustration due to the hole motion. \cite{SS}
Such a state would not co-exist with charge order, as can
 be shown in the context of effective Landau theory. \cite{Zachar}
However  it is still possible that in the presence of anisotropic  Dzyaloshinski-Moriya 
 interactions the spin spiral could cause  weak lattice modulation  with period generically
half that of the spiral.  
While the spiral ground  state  has a  tendency to be unstable toward phase separation, 
 we have  shown recently that in the extended  $t-t'-t''-J$ model the spiral
 can be stabilized by the presence of the additional hoppings. \cite{SK} 
 The spiral description was successfully applied  to explain
  magnetic properties of LSCO, such as the location of the
 elastic neutron scattering peaks and  the change
of the incommensurability direction by 45$^{\circ}$ across the
 superconductor-insulator boundary ($x = 0.055$). \cite{SK1} 
 The spiral  was also proposed as a candidate for the ground state
 in the spin-glass (insulating) phase for $0.02 < x < 0.055$. \cite{HCS,Juricic,SK1}

In the present work we  address transport properties within the spiral
framework. Our main motivation comes from the recent experimental
data in the spin-glass phase of  LSCO showing transport
anisotropies as large as 50\% to 70\%, both in  DC and AC measurements.  \cite{Ando, Dumm}
 It has been argued that these data provide indirect support
 to the notion of fluctuating stripes or electronic liquid crystals, but
no quantitative theory exists  that takes  these concepts into account. \cite{Kivelson1} 
 Recent infrared experiments that carefully identify phonon modes in LSCO
 once again find no charge ordering tendencies in the spin-glass phase of this
 material. \cite{Padilla} Therefore we  take the  point of view that  
 the ground state has a spiral spin structure, and calculate the
 transport anisotropy in the variable-range hopping (VRH) regime 
for low temperature and frequency, where the anisotropy  has the largest value.
 We show that the spatial anisotropy of the hole wave-function in a spiral
 translates into anisotropy of the hopping transport. Our main result is that
for microscopic parameters appropriate for LSCO (within the $t-t'-t''-J$ model),
 the magnitude of the anisotropy  is large (40\%-90\%, depending on temperature),
 and close to the one found experimentally. 
 Thus we demonstrate that  
the  transport anisotropy data can be explained  {\it quantitatively} within the spiral theory
 which does not involve any tendency of the holes to self-organize into
charge stripes. The anisotropy was also analyzed in Ref.~\onlinecite{Juricic} 
on the basis of topological defect scattering in a spiral; however the results are
applicable to the quasi-metallic (higher temperature) regime where the anisotropy
is very small, of the order of several percent. We emphasize that in the present
work we  consider the low-temperature, strongly-localized regime where  it is clear from experiment
 that the anisotropy is the largest  (10-20 times larger than the
 value in the metallic regime). Finally we mention that the  problem was
also addressed within the spin-charge separation scenario which provides
 a description of the pseudogap phase \cite{Marchetti} and could lead to effective ``insulating"
behavior; however within this approach the transport anisotropy is linked to the
 magnetic correlation length anisotropy which can be taken from experiment but
 cannot be explicitly calculated.

\subsection{Summary of Previous Results and Notation}

Our starting point  is the description of the spin-glass phase ($0.02 < x < 0.055$)
 developed in Ref.~\onlinecite{SK1} which is briefly summarized below.
 First, a single hole resides near the points  ${\bf k}_0=(\pm\pi/2,\pm\pi/2)$,
 with a quadratic dispersion around them 
$\epsilon_{\bf k}\approx \frac{\beta_1}{2}k_1^2+\frac{\beta_2}{2}k_2^2$,
 where $k_1$ is perpendicular
 to the face of the magnetic Brillouin zone and  $k_2$ is parallel to the face.
 Within the effective $t-t'-t''-J$ model for LSCO, the parameters are taken to
 be: $t/J=3.1, t'/J \approx -0.5, t''/J \approx 0.3$, where $J \approx 125 \ \mbox{meV}$.
 From now on we measure energies in units of $J$ (i.e. set $J=1$) and lengths
 in units of the lattice spacing $a$ (we set $a=1$).
By using the self-consistent Born approximation
one finds that these parameters lead to an almost isotropic dispersion $\beta_1 \approx \beta_2 = \beta
\approx 2.2$, and a quasiparticle residue around the nodal points $Z \approx 0.34$.
 A  hole trapped by the Coulomb field of the Sr ion
generates a spiral distortion of the N\'eel background,
 which can be parametrized as:
$|\mbox{i}\rangle = \!  e^{i\theta({\bf{r_i}}){{\bf{m}}}
\cdot{\bf{\sigma}}/2}| \! \uparrow\rangle,
 |\mbox{j}\rangle =  \! e^{i\theta({\bf{r_j}}){\bf{m}}
\cdot{\bf{\sigma}}/2}| \! \downarrow\rangle$,
$\mbox{ i $\in$  ``up''  sublattice}, \mbox{   j $\in$  ``down''  sublattice}$,
where ${\bf m}$ is an arbitrary unit vector perpendicular to the
 quantization axis of the states $|\uparrow\rangle$, $|\downarrow\rangle$.
The angle $\theta({\bf r})$, measuring deviations from
 collinearity is given by: \cite{SK1}
\begin{equation}
\label{dipole}
\theta({\bf r})=\frac{Zt}{\sqrt{2}\pi\rho_s}\frac{({\bf e_{\pm}\cdot r})}{r^2}
\left[1-e^{-2\kappa r}(1+2\kappa r)\right], 
\end{equation}
where  $\rho_s \approx 0.18$ is the spin stiffness, and $1/\kappa\sim 3-5$ is the localization
length of the orbital part of the wave function. The value of $1/\kappa$ is
 extracted from experimental data, \cite{SK1} and 
is generally expected to increase with doping. The unit vector 
${\bf e}_{\pm} = \frac{1}{\sqrt{2}} (1,\pm 1)$, in the usual square-lattice
 coordinate system.
At finite doping ($0.02 < x < 0.055$)  the interaction of the long-range dipole distortions 
from holes trapped at different Sr ions leads to spiral magnetic  order, \cite{SK1}
characterized by average $\bar{\theta} \propto {\bf e_{\pm}\cdot r}$, as
shown in Fig.~\ref{Fig1}(a).
In a perfect square lattice the spiral can be directed along any diagonal $(1,\pm1)$.
However from the location  of the elastic neutron scattering peaks \cite{Wakimoto} 
 the   incommensurability is  determined to be  along the orthorhombic $\hat{b}$ direction.
In our picture this means that the orthorhombic deformation pins the direction
 of the spiral, as shown in Fig.~\ref{Fig1}, and  thus we set 
${\bf e} = {\bf e}_+ = \frac{1}{\sqrt{2}} (1,1)$. In what
follows the exact nature of the pinning mechanism, to be discussed elsewhere,
 will not be important.

Experimentally the magnetic correlation length is
 finite $\xi \sim 6-20$ (decreases with increasing doping). \cite{Wakimoto,Fujita}
On the theoretical side it was argued that $\xi$ is finite
 due to topological defects that lead to frustration of the long-range
spiral order. \cite{HCS,SK1}
Since the localization length is less than the  magnetic scale,  
$1/\kappa < \xi$, we expect  that the effective description of transport
 properties, developed below in terms of
the  one-hole wave function, is quantitatively valid
 also at finite (small) doping (i.e. the topological
 frustration mechanism  does not affect our considerations). 
 It should be noted that the above inequality 
 while being  explicit is also the strongest (most restrictive) condition we could give. Its refinement
 would have to come out of a  detailed theory of IC magnetism in the spin-glass
 phase (i.e. a self-consistent theory that takes into account both the effective disorder
 and the spiral formation, generated by the doped holes).
Purely theoretical arguments aside, transport measurements in the
 doping range  0\%-4\% suggest that the system is in a strongly-localized
regime at low temperatures, and we thus  expect our calculations to be valid
 as long as doping is not too close to the insulator-metal boundary (at 5.5\%).

The rest of the paper is organized as follows. In Section II 
we analyze the properties of a localized hole in a spiral background, and
in Section III we use the hole's wave function  to
 calculate the in-plane transport anisotropies in the variable-range
hopping regime. Section IV contains our conclusions.

\section{Localized hole in the presence of magnetic spiral correlations} 
The coupling $H_{SP}$
between the spin of the magnetic background (angle $\theta({\bf r})$) and the orbital
 wave-function of the hole $\chi({\bf r})$  generates the spiral and
 has the form: \cite{SK1}
%$H_{SP} = - \sqrt{2} Z t \int ({\bf e}\cdot{\bf \nabla}\theta)\chi^{2}({\bf r}) d^{2}r$.
\begin{equation}
H_{SP} = - \sqrt{2} Z t \int ({\bf e}\cdot{\bf \nabla}\theta)\chi^{2}({\bf r}) d^{2}r \ .
\end{equation}
The effective Schr\"{o}dinger equation is then:
\begin{equation}
\label{Sch}
\left(-\frac{\beta}{2}\nabla^2 -\frac{q^2}{r} -\sqrt{2}Z t   ({\bf e}\cdot{\bf \nabla}\theta)
\right)\chi({\bf r})=\epsilon\chi({\bf r}) \ .
\end{equation}
Here the Coulomb potential of the $Sr$ ion which keeps
 the hole localized is $\frac{q^2}{r} = \frac{q_{0}^2}{{\cal{E}}_{e} r}$,
 where $q_{0}$ is the unit charge and ${\cal{E}}_{e}$ is the effective dielectric constant
 known to be quite large in the copper oxides, ${\cal{E}}_{e} \sim 30-100$
(increases with doping). \cite{Kastner}
\vspace{.6cm}
\begin{figure}[ht]
\centering
\includegraphics[height=145pt,keepaspectratio=true]{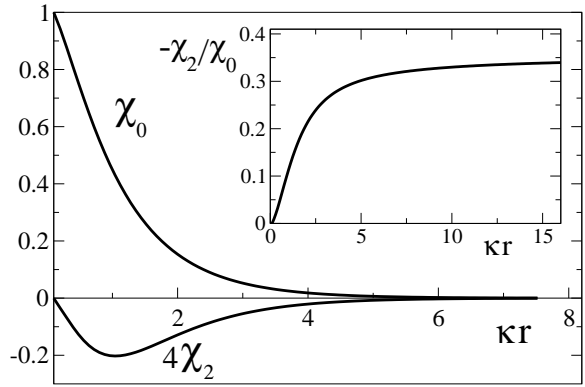}
\vspace{.4cm}
\caption{Solution of (\ref{Sch}) in terms of the expansion (\ref{solution}),
 for $\Lambda = 1.$
Only the first two (largest) components are shown.
The normalization of $\chi_0$ is arbitrary. Inset: The ratio
$-\chi_{2}/\chi_{0}$. } 
\label{Fig2}
\end{figure}
The last term in (\ref{Sch}) determines the anisotropy of $\chi({\bf r})$  and
 is given by
\begin{eqnarray}
&&\sqrt{2}Z t   ({\bf e}\cdot{\bf \nabla}\theta) = \frac{Z^2t^2}{\pi\rho_s}\frac{1}{r^{2}}
[ f(\kappa r) - g(\kappa r)  \cos(2\varphi) ] \ , \nonumber \\
&& g(z) = 1 -  ( 1 + 2z + 2 z^{2} ) e^{-2z}, \  f(z) = 2 z^{2} e^{-2z}. 
\end{eqnarray}
The coordinate system is chosen so that ${\bf e}$ is parallel to the $\hat{b}$ axis
and $\varphi$ is the polar angle (Fig.~\ref{Fig1}).
The value of $\kappa$ (and consequently $\epsilon$)  can be 
found  by a variational minimization of
the total energy: \cite{SK1}
\begin{equation}
\label{Lambda}
\kappa = \frac{2 q^{2}/ \beta}{1 - \frac{\Lambda}{2}},  \ \epsilon = - \beta \kappa^{2}/2, 
\  \ 
 \Lambda \equiv \frac{Z^2t^2}{\pi \beta \rho_s} \ .
\end{equation}
We will use  $\kappa \sim 0.3-0.4$, consistent with
 experiment, \cite{SK1} rather than rely on (\ref{Lambda}) which
 contains the uncertainty related to the exact value of the
 dielectric constant.
The dimensionless parameter $\Lambda$, as defined in (\ref{Lambda}), 
characterizes the coupling between the orbital motion of the hole and the
 deformation of the spin background. Substituting the values of  $Z, \beta$
 appropriate for  LSCO (discussed in Section I.A), we obtain $\Lambda \approx 1$,
 which will be used from now on.

Let us look for solution of  (\ref{Sch}) in terms of the series in angular harmonics
\begin{equation}
\label{solution}
\chi(\kappa r,\varphi)=\chi_{0}(\kappa r) + \chi_{2}(\kappa r) \cos(2\varphi)  + 
\chi_{4}(\kappa r) \cos(4\varphi) + \dots
\end{equation}
Then Eq.~(\ref{Sch}) leads to a set of coupled equations for $\chi_{i}, i=0,2,4,..$.
We have  solved  these equations numerically  and the results for the first two 
harmonics are presented in Fig.~\ref{Fig2}. The next harmonics are found to be  small, for
example $\chi_{4}/\chi_{0} < 2 \times 10^{-2}$, and can be in fact safely neglected. 
At large distances all the wave-functions
behave as  $\chi_i \propto e^{-\kappa r}$, while at finite distances
this behavior is substantially modified, as shown in  Fig.~\ref{Fig2}.

\section{Transport anisotropies in the variable-range hopping (VRH) regime} 
\subsection{In-plane DC Resistivity Anisotropy}
Below  characteristic temperature $T_{VRH} \approx 20-30 \ \mbox{K}$,
the low-temperature resistivity of LSCO in the spin-glass phase can be described by
 the 2D version of the  Mott  VRH formula  $\rho \sim \exp{(T_0/T)}^{1/3}$.  
\cite{Keimer,Kastner,Lai,Suzuki}  Here $T_0$ depends somewhat on doping and sample quality and
generally decreases when doping increases (and thus conduction becomes easier).  
As an estimate, for  example at 4\% doping the data of Ref.~\onlinecite{Keimer} are well
 fit with $T_0 \approx 500 \ \mbox{K}$; \cite{Lai} analyzing the curves of   Ref.~\onlinecite{Suzuki}
 we obtain $T_0 \approx 200-300 \ \mbox{K}$, and from Ref.~\onlinecite{Ando} we 
have extracted  $T_0 \approx 200 \ \mbox{K}$.

Within the spiral framework the
 physical idea behind the resistivity anisotropy is quite simple.
 According to (\ref{solution}) the  hole wave function  acquires an elliptic 
deformation induced by the
spin spiral, as shown schematically in Fig.~\ref{Fig1}(b). Hence  
hopping in the $\hat{b}$-direction is less probable than hopping in the $\hat{a}$-direction.
Using (\ref{solution}), and essentially following the derivation
 of  Mott's  result, \cite{Mott} we obtain for the resistivity anisotropy: 
%\begin{equation}
%\label{DCA}
%\frac{\rho_b(T)}{\rho_a(T)} =
% \frac{\int_{0}^{2 \pi} \sin^{2}(\varphi)\Phi^{2}(\kappa r_T,\varphi)
% d\varphi}{\int_{0}^{2 \pi} \cos^{2}(\varphi)\Phi^{2}(\kappa r_T,\varphi) d\varphi}, \ 
% \Phi(\kappa r,\varphi) \equiv \frac{\chi(\kappa r,\varphi)}{\chi_{0}(\kappa r)}
%\end{equation}
\begin{eqnarray}
\label{DCA}
&&\frac{\rho_b(T)}{\rho_a(T)} = \frac{\int_{0}^{2 \pi} \sin^{2}(\varphi)\Phi^{2}(\kappa r_T,\varphi)
 d\varphi}{\int_{0}^{2 \pi} \cos^{2}(\varphi)\Phi^{2}(\kappa r_T,\varphi) d\varphi},\\
&&\Phi(\kappa r,\varphi) \equiv \frac{\chi(\kappa r,\varphi)}{\chi_{0}(\kappa r)} \ . \nonumber
\end{eqnarray}
Here $r_{T}= \frac{1}{2 \kappa} (T_0/T)^{1/3}$ is the VRH 
length, \cite{Mott} and in order for the approach to be justified we must certainly have 
$\kappa r_T > 1$. Eq.~(\ref{DCA})  reflects the difference in the wave-function
overlap for the two electric field directions. The factors $\sin^{2}(\varphi),\cos^{2}(\varphi)$ arise
from the fact that the conductivity varies with the square of the  carrier
 jump distance projection in the electric field direction.  \cite{Mott}
% $R={\bf R}.{\bf E}$ 
%  (${\bf E}$ is the field). \cite{Mott}
 Since the hole's dispersion
 is isotropic ($\beta_{1}=\beta_{2}=\beta$), the exponential part of the wave function
 is isotropic and consequently the resistivity anisotropy is expected to 
arise from the anisotropy of the exponential prefactors, whose overlap 
leads to (\ref{DCA}). Both the VRH length and $T_0$ thus remain
isotropic within this formulation.

\begin{figure}[t]
\centering
\includegraphics[height=165pt,keepaspectratio=true]{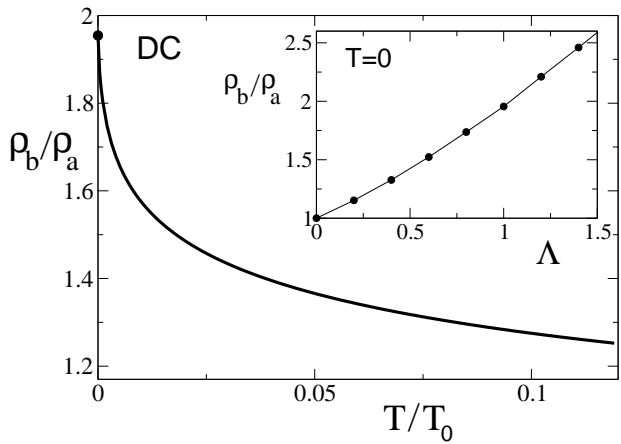}
\caption{In-plane DC resistivity anisotropy in the VRH regime, for $\Lambda=1$.
 Inset: Maximum ($T=0$) anisotropy versus  $\Lambda$. }
\label{Fig3}
\end{figure}

It is clear from (\ref{DCA}) that  as $T$ decreases (i.e. $\kappa r_{T}$
increases) the anisotropy grows, due to the increase of $|\chi_{2}/\chi_{0}|$
 (Fig.~\ref{Fig2}(Inset)) and consequently the more pronounced angular dependence
 of $\Phi(\kappa r_{T},\varphi)$. The results are summarized
 in Fig.~\ref{Fig3}, where also the evolution of
the zero temperature anisotropy $\rho_b(T=0)/\rho_a(T=0)$ as a function of $\Lambda$
is shown  for completeness in the inset.
At very low temperature ($T \lesssim 1 \ \mbox{K}$, \cite{Lai} i.e. $T/T_{0} \lesssim 10^{-3}$)
 we would have to take into account a crossover to the Coulomb gap regime,
 which however would practically  not  influence  the curve in  Fig.~\ref{Fig3}.

 The data of Ref.~\onlinecite{Ando}
were taken at temperatures $T > 10 \ \mbox{K}$, meaning that the lowest
 ratio $T/T_{0} \sim 0.05$ (we take $T_{0} \sim 200 \ \mbox{K}$).
As the  temperature  increases beyond  $T_{VRH} \approx 20-30  \ \mbox{K}$ when
$\kappa r_{T} \approx 1$, the approach  based on VRH conduction ceases to be valid,
 as the conduction mechanism changes to impurity band conduction
 and eventually quasi-metallic behavior.  
In the most relevant low-temperature range
  below $T_{VRH}$ (corresponding to largest anisotropy),
 both the calculated magnitude of $\rho_b/\rho_a$ and its
 temperature dependence are very close to the experimental
 results, \cite{Ando} although in these experiments the temperature
 is not low enough to penetrate the ``deep'' VRH regime $T/T_{0} \lesssim 0.05$
 where the anisotropy should increase even further. 
It should be also noted that there exists quite a bit of uncertainty
in the determination of $T_{0}$ and hence in the determination of
the exact value of $\rho_b/\rho_a$ from Fig.~\ref{Fig3}.

\vspace{0.7cm}

\subsection{AC Resistivity Anisotropy} 
At finite
frequency  and temperature the calculation of the AC conductivity
 is a very complicated problem. However in the ``quantum'' limit
 when the frequency $\omega \gg T$, the VRH AC conduction is expected to
 be dominated by resonant absorption by singly occupied pairs (without
 involvement of phonons), and is usually relevant in doped semiconductors. \cite{Efros}
%A finite density of states at the chemical potential is necessary for this mechanism
% to be operational, and indeed underdoped LSCO in the spin-glass phase
% behaves as an Anderson insulator with ``arcs'' of Fermi surface near
% the nodal points  and states exist in what was 
% the charge-transfer gap at zero doping \cite{Yoshida}.
 The VRH length in the AC regime 
(neglecting for a moment the angular dependence of the states) 
 is logarithmic: \cite{Efros}  $r_{\omega} = (1/\kappa) \ln(2|\epsilon|/\omega)$,
which is the main difference from the DC case.
  From now on we denote 
$\Omega \equiv \omega/(2|\epsilon|)$.
The formula for  $r_{\omega}$ follows from the fact that upon evaluation
 of the conductivity, the most effective pairs are the ones that satisfy: \cite {Efros}
$\omega = 2 I(r)$, where $I(r)=I_{0}e^{-\kappa r}$ is the overlap integral,
 and a typical estimate of the prefactor is $I_{0} \approx |\epsilon|$.
The functional dependence of the conductivity in 2D  at low frequency $\ln(1/\Omega) \gg 1 $,
 in the leading logarithmic approximation, is
given by the Mott-Shklovskii-Efros expression: $\sigma(\omega) \sim \omega r^{3}_{\omega} [ \omega
+ q^{2}/r_{\omega} ]$, where the second term takes into account the Coulomb interaction
 in the resonant pair. \cite{Efros}

\begin{figure}[ht]
\centering
\includegraphics[height=150pt,keepaspectratio=true]{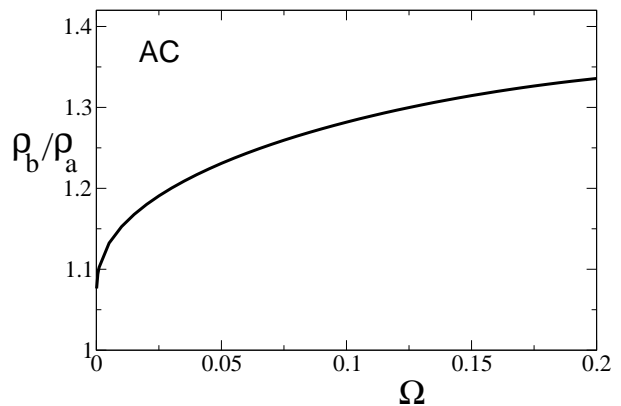}
\caption{AC VRH resistivity anisotropy versus $\Omega=\omega/(2|\epsilon|)$
 in the regime $T \ll \omega$, and  $\ln(1/\Omega) \gg 1$, for $\Lambda=1$.}
\label{Fig4}
\end{figure}

The generalization of the  Mott-Shklovskii-Efros formula to the anisotropic case is straightforward,
 as it amounts to taking into account the non-exponential (angle-dependent) part of the wave-function,
leading to the 
 replacement (with logarithmic accuracy):
 $\ln(1/\Omega) \rightarrow \ln\left(\Phi(\kappa r_{\omega},\varphi)/\Omega\right)
\equiv L_{\omega,\varphi}$,
where $\Phi$ is defined in (\ref{DCA}).
% We have generalized these results to the case  when the
%localized wave-functions  are given by (\ref{solution}). For the
Taking also into  account  the expressions for $\epsilon, \kappa$ (\ref{Lambda}),
 we obtain for the resistivity anisotropy:     
\begin{eqnarray}
\label{ACA}
&&\frac{\rho_b(\omega)}{\rho_a(\omega)} = \frac{\int_{0}^{2 \pi} \sin^{2}(\varphi)
{\cal F } (\omega,\varphi)
 d\varphi}{\int_{0}^{2 \pi} \cos^{2}(\varphi)
{\cal F } (\omega,\varphi)
d\varphi}, \\
&&{\cal F } (\omega,\varphi) =  L_{\omega,\varphi}^2\left(1+\frac{2\Omega}{1 - 
\Lambda/2} L_{\omega,\varphi}\right) \ . \nonumber
\end{eqnarray}
%where
%\begin{equation}
%{\cal F } (\omega,\varphi)=L^2+
%\ln^{2}\left(\frac{\Phi(\kappa r_{\omega},\varphi)}{\Omega}\right) +  
%\frac{2\Omega}{1 - \frac{\Lambda}{2}}\ln^{3}\left(\frac{\Phi(\kappa r_{\omega},\varphi)}{\Omega}\right).
%\frac{2\Omega}{1 - \Lambda/2}L^3 \ .
%\end{equation}
The expression (\ref{ACA}) is plotted in Fig.~\ref{Fig4}.
Due to the logarithmic dependence, in the theoretical limit of
 zero frequency the anisotropy would vanish (albeit very slowly),
 i.e. $\rho_b(\omega=0)/\rho_a(\omega=0)=1$. However one must keep
 in mind that the above expressions are not valid at
 arbitrary low frequencies, as we must have $\omega > T$.
Generally we find that the  (maximum) AC anisotropy of 30-40\% is somewhat smaller than
 the DC anisotropy.

\vspace{.7cm}

When attempting to compare our results to the
 experimental data of Ref.~\onlinecite{Dumm} we  realize 
that $\omega$ and $T$ are not sufficiently low, nor
 is the difference between them sufficiently high to
 justify the separation into ``quantum" and ``thermal"
VRH conduction and consequently the Mott-Shklovskii-Efros approach.
The temperature used is $T=13 \ \mbox{K}$ and at  the lowest frequency
 of $\omega = 20  \ \mbox{cm}^{-1} \approx 2.5 \ \mbox{meV}$, estimating
 for the hole energy $|\epsilon| = \beta \kappa^{2}/2 \approx 5-10 \ \mbox{meV}$,
 we have $\Omega \sim 0.13-0.25$.
Moreover, upon increasing the frequency to $80-100 \ \mbox{cm}^{-1}$,
 a change of behavior from insulating to conducting takes place,
a broad peak develops, \cite{Dumm}  and consequently it is not clear
 to us that the AC data are ever in a clean VRH regime.
In spite of all this the magnitude of the calculated  anisotropy
 agrees reasonably well with experiment.
It should be experimentally possible  to lower the temperature (towards the mK range) as well 
as $\omega$ in order to probe the VRH AC anisotropy more reliably and
compare with our theory. 

\section{Discussion and Conclusions}
We would like to reiterate  that in our theory there is no tendency of the holes to 
form charge stripe-like structures. Nevertheless the resistivity
shows anisotropic behavior  (and so does the  uniform magnetic susceptibility,
to be discussed separately).
It is particularly important, in our view, that the calculations presented
 in this work produce quantitative results,  and we would thus hope that
 the successes of the present formulation would stimulate further
 exploration of the spiral and similar scenarios, not involving
 any charge-ordering tendencies. The present work is relevant to the
spin-glass regime of La$_{2-x}$Sr$_x$CuO$_4$, $0.02 < x < 0.055$.
However  we have also argued  \cite{SK,SK1} that the spin spiral structure
could exist in the superconducting state $x >0.055$. The theory naturally
explains why the incommensurate direction, determined from elastic neutron
scattering,  rotates by 45$^\circ$ exactly at
the insulator-superconductor transition point.
On the other hand it is usually  argued that at  $x=1/8$ both in charge-ordered LNSCO \cite{Tranquada}  
and LBCO, \cite{Hucker}  non-collinear spiral order is not consistent with
experiment. We would indeed not claim that 
the spiral ground state is stable at that particular doping
 since in fact the  spiral becomes commensurate with the lattice
and therefore  significant  changes in the ground state could 
occur due to the spin-lattice coupling.

In conclusion, starting from a spiral ground state which
unambiguously follows from the extended $t-J$ model, we have calculated the 
in-plane anisotropy of the low-temperature DC and low-frequency AC resistivity 
of La$_{2-x}$Sr$_x$CuO$_4$ in the spin-glass phase. The theory has no fitting 
parameters and the calculations are performed for the variable-range hopping 
regime.
 Within the spiral description the transport anisotropy is due to
the anisotropy of the hole wave-function since the hole hops in
 a non-collinear (spiral) spin background. 
 The AC anisotropy reasonably agrees with experiment in spite of the
fact that the data are on the border of the VRH regime.
The experimental data for the  DC resistivity are  well within the VRH regime and here
both the calculated  magnitude and  temperature dependence of the  anisotropy agree with experiment 
very well.

\begin{acknowledgments}
We are grateful to  L. Benfatto, J. Haase,  and  
 D. Poilblanc  for valuable discussions and comments. 
V.N.K. acknowledges the   support of the  Swiss National Fund.
\end{acknowledgments}

\end{document}